# Energy Bands and Breakdown Characteristics in Al$_2$O$_3$/UWBG AlGaN Heterostructures

Seungheon Shin[1,a], Kyle Liddy[1], Yinxuan Zhu[1], Chandan Joishi[1], Brianna A. Klein[3], Andrew Armstrong[3], Andrew A. Allerman[3], Siddharth Rajan[1,2,a]

[1]*Department of Electrical and Computer Engineering, The Ohio State University, Columbus, Ohio 43210, USA*
[2]*Department of Materials Science and Engineering, The Ohio State University, Columbus, Ohio 43210, USA*
[3]*Sandia National Laboratories, Albuquerque, New Mexico 87123, USA.*

**Abstract:**

We report on energy bands and breakdown characteristics of Al$_2$O$_3$ dielectrics on ultra-wide bandgap (UWBG) AlGaN heterostructures. Metal-dielectric-semiconductor structures are important to sustain high fields needed for future high-performance UWBG transistors. Using systematic experiments, we determined the fixed charge density (> $10^{13}$ cm$^{-2}$) the dielectric/interface, and electric fields in the oxide of under flat-band conditions in the semiconductor. Low gate-to-drain leakage current of up to $5 \times 10^{-7}$ A/cm$^2$ were obtained in the metal-oxide-semiconductor structures. In lateral metal-semiconductor-insulator test structures, breakdown voltage exceeding 1 kV was obtained with a channel sheet charge density of $1.27 \times 10^{13}$ cm$^{-2}$. The effective peak electric field and average breakdown field were estimated to be > 4.27 MV/cm and 1.99 MV/cm, respectively. These findings demonstrate the potential of Al$_2$O$_3$ integration for enhancing the breakdown performance of UWBG AlGaN HEMTs.

[a] Authors to whom correspondence should be addressed

Electronic mail: *shin.928@osu.edu, rajan.21@osu.edu*

Ultra-wide bandgap (UWBG) materials are emerging as promising future semiconductors, offering a pathway toward terahertz switching frequency and high-voltage applications. Among UWBG materials, AlGaN has garnered significant attention due to its potential to achieve a higher Johnson's figure of merit (JFOM), a key metric evaluating both on-state and off-state performance by incorporating cut-off frequency ($f_T$) and breakdown voltage ($V_{BR}$). UWBG AlGaN offers distinct advantages based on its high saturation velocity ($\sim 2 \times 10^7$ cm/s) despite relatively low mobility ($\sim 250$ cm$^2$/V·s), and a wide bandgap energy ($\sim 6.2$ eV) [1-4], which contributes to a high critical field ($\sim 12$ MV/cm) [5-6]. However, the realization of the theoretical JFOM limit of UWBG AlGaN ($\sim 22 \times 10^{12}$ Hz·V) [1] is hindered by challenges such as high contact resistance and premature breakdown caused by excessive gate leakage. To address the contact resistance issue, various approaches have been explored, and Zhu et al. recently demonstrated a record-low $R_C$ of 0.25 Ω·mm [7]. Maximizing JFOM requires the simultaneous reduction of sheet resistance ($R_{SH}$) and contact resistance ($R_C$) to enhance $f_T$, as well as effective field management strategies to handle extreme electric fields resulting in improving $V_{BR}$. To prevent breakdown at the gate electrode, gate dielectric integration—using materials such as $SiO_2$, $SiN_x$, AlN, and $Al_2O_3$—has been extensively studied in AlGaN/GaN and GaN-based devices [8-29]. Aluminum oxide has been shown to be an excellent choice for device applications due to widespread adoption and high-quality thin layer via plasma-enhanced atomic layer deposition (PEALD) on III-nitride devices [28, 29]. Previous reports have investigated the properties of $Al_2O_3$ on GaN and low-composition AlGaN [36-38], including evidence for positive fixed interface charges ($\sim 1.9 \times 10^{13}$ cm$^{-2}$) at $Al_2O_3$/GaN interfaces [30, 39]. However, reports on the interfacial characteristics between dielectrics and UWBG AlGaN remain limited, leaving critical questions about interface charges and leakage suppression mechanisms unresolved. In this work, we investigate the interfacial and breakdown characteristics of PEALD $Al_2O_3$ integrated on UWBG AlGaN HEMT structures of varying Al compositions.

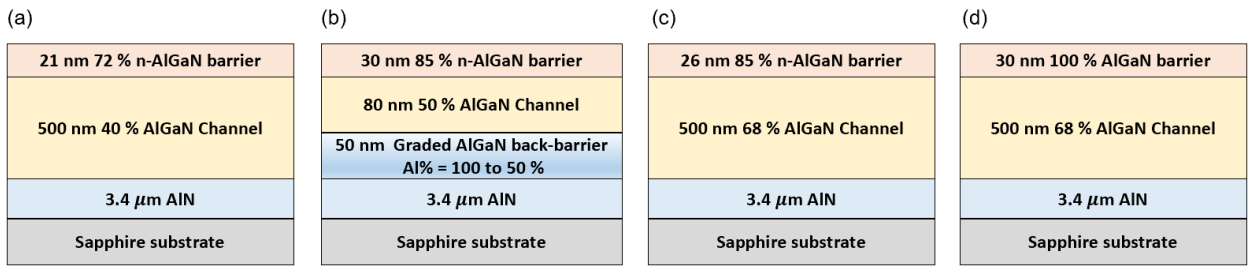

Figure 1. Schematics of used epitaxy structures (a) 72/40, (b) 85/50, (c) 85/68, (d) 100/68

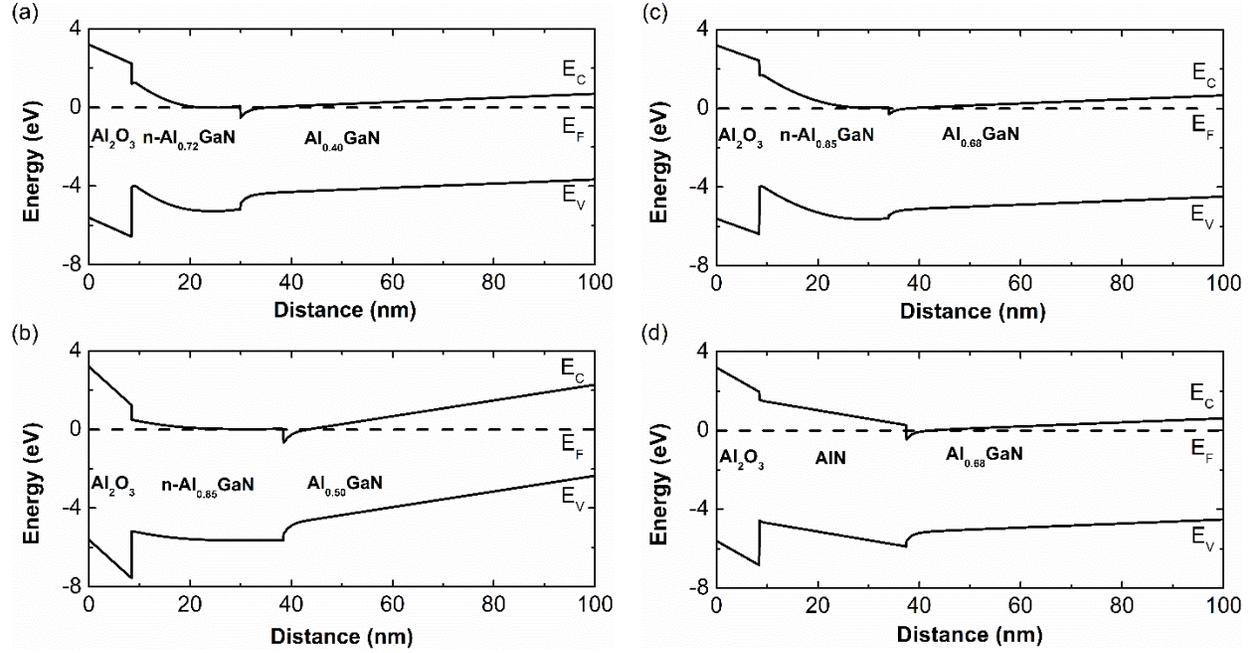

Figure 2. Calculated energy band diagram in equilibrium under the gate regions of (a) 72/40, (b) 85/50, (c) 85/68, (d) 100/68

The epitaxial structures used in this study were grown on sapphire substrates via a TNSC-4000HT metal-organic chemical vapor deposition (MOCVD) reactor on pre-grown AlN/Sapphire templates. Figure 1 shows the four different $Al_xGa_{1-x}N/Al_yGa_{1-y}N$ HEMT structures (denoted as x/y for each sample). For each epitaxial structure, the calculated energy band diagrams including 8.5 nm of $Al_2O_3$ under the gate region were shown in Figure 2. Assumed barrier height for Ni/$Al_2O_3$ interface is 3.2 eV [34-35]. Hall effect measurements for the 85/50 sample exhibited a sheet resistance of 1.69 kΩ/□, a sheet charge density of $1.92 \times 10^{13}$ cm$^{-2}$, and an electron mobility of 192 cm²/V·s. Non-ideal ohmic contacts precluded Hall measurements on the other samples, but CV measurements (Table 1) provide the relevant information on the sheet charge density.

Table 1. Estimated sheet charge density with C-V measurements

|  | 72/40 | 85/50 | 85/68 | 100/68 |
| --- | --- | --- | --- | --- |
| Sheet charge density (cm$^{-2}$) | 0.86 | 1.38 | 0.8 | 1.27 |

For device processing, Zr-based ohmic contacts (Zr/Al/Mo/Au = 15/120/40/50 nm) were deposited via E-beam evaporation and annealed at 950 °C. Following a buffered oxide etch (BOE) surface treatment, an $Al_2O_3$ layer was deposited on each sample via plasma-enhanced atomic layer deposition (PEALD) using a trimethylaluminum metal-organic precursor at a substrate temperature of 250 °C, targeting a 30 nm thickness. After the $Al_2O_3$ deposition, three different $Al_2O_3$ thicknesses ($t_{ox}$ = 8.5, 13, and 28 nm) were prepared via BOE wet etching. Subsequently, Ni/Au/Ni (20/120/20 nm) metal stacks were evaporated to form the gate metal. The dielectric constant and refractive index of the deposited $Al_2O_3$ were estimated to be 9 (from C-V measurements) and 1.73 (from spectroscopic ellipsometry).

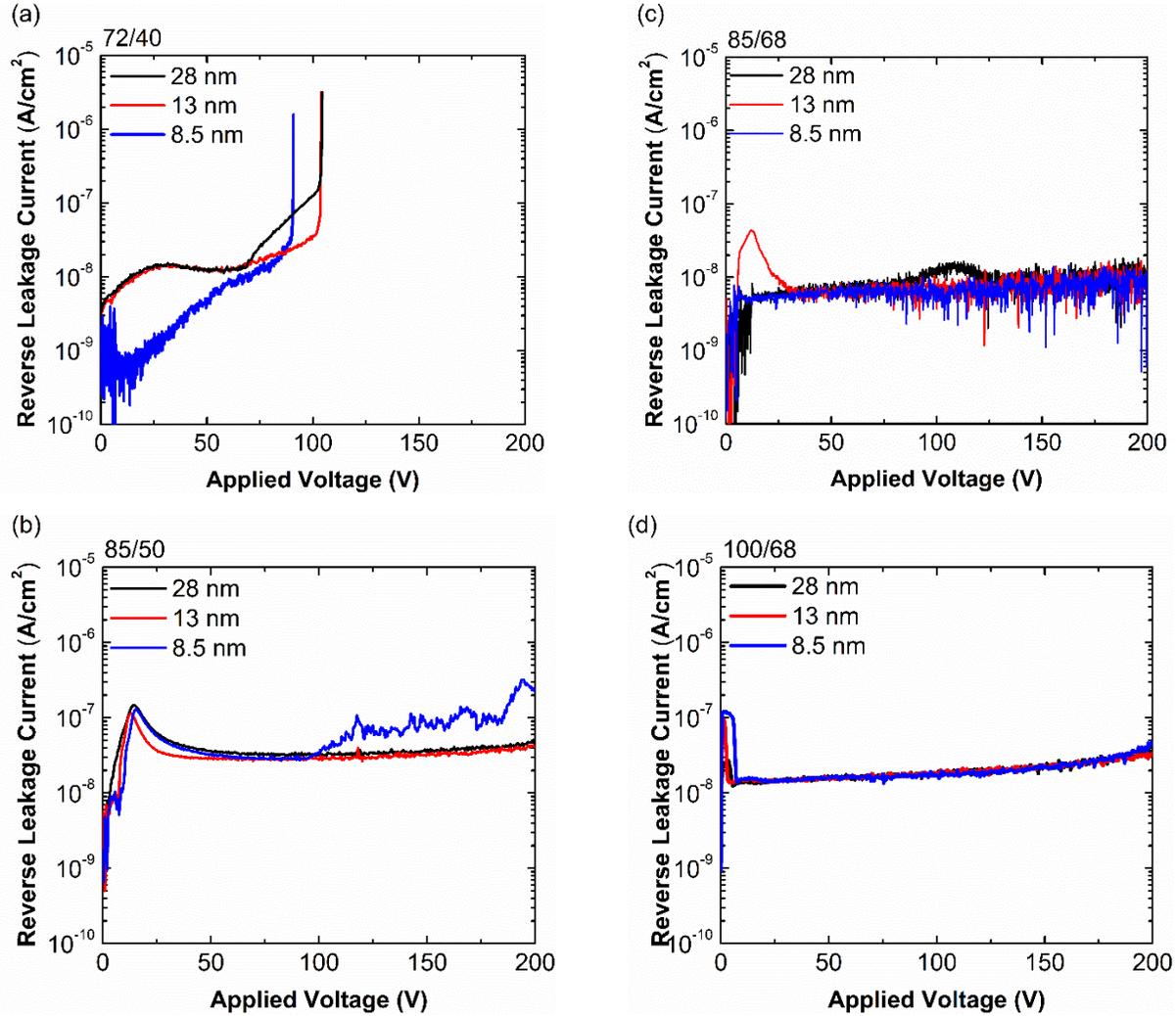

Figure 3. two terminal leakage current measurements (a) 72/40, (b) 85/50, (c) 85/68, (d) 100/68

To investigate the leakage current characteristics, two-terminal reverse I-V measurements were performed on all four samples using Keysight B1500A. All samples exhibited low leakage current (~ 5 × $10^{-7}$ A/cm$^2$) up to 200 V or device breakdown with a 5 μm gate-to-drain spacing (Figure 3). These results indicate that the PEALD $Al_2O_3$ layer effectively suppresses gate-to-drain leakage compared to previous UWBG AlGaN HEMT demonstrations, which incorporated Schottky gate structures and $SiO_2$ integrated MOSHFET [31, 32]. Furthermore, the gate-to-drain leakage characteristics remained consistently low, regardless of $Al_2O_3$ thickness.

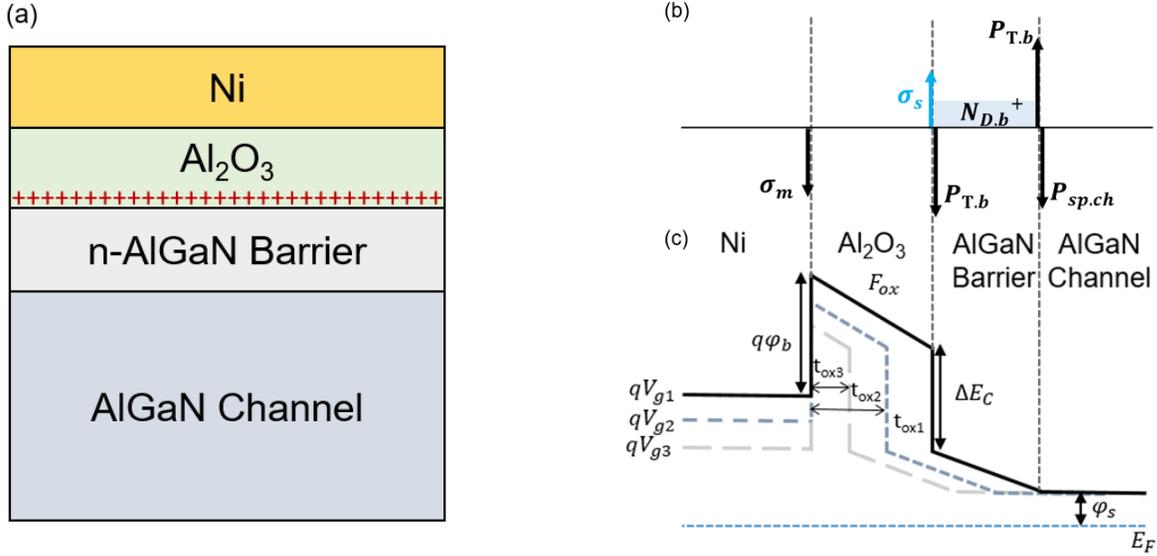

Figure 4. (a) Schematic of Ni/Al$_2$O$_3$/Al$_x$Ga$_{1-x}$N/Al$_y$Ga$_{1-y}$N structure, Flat-band condition Ni/Al$_2$O$_3$/Al$_x$Ga$_{1-x}$N/Al$_y$Ga$_{1-y}$N (b) charge diagram, (c) energy band diagram

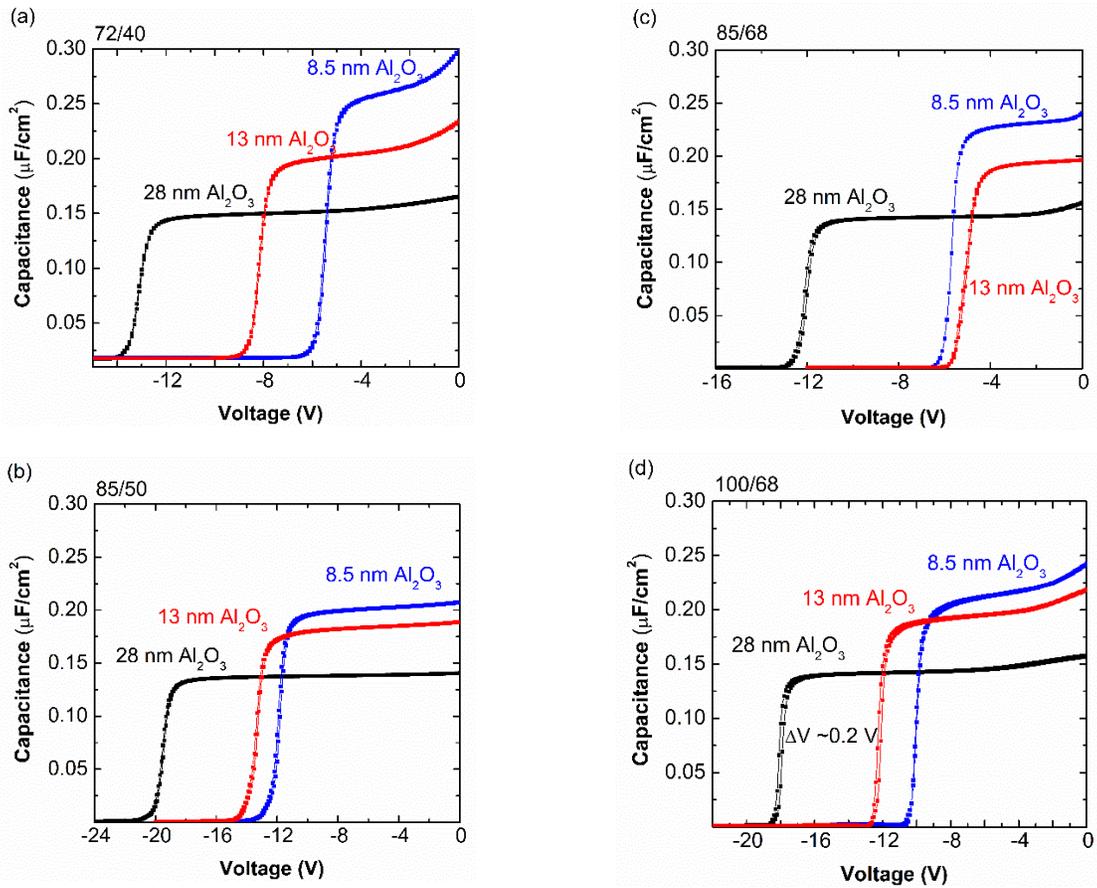

Figure 5. C-V measurement results for each Al$_2$O$_3$ thickness (a) 72/40, (b) 85/50, (c) 85/68, (d) 100/68

To explore the fixed charge and internal oxide field intensity, a modified C-V method was employed, building upon prior studies of Al₂O₃/GaN interfacial characteristics [30]. Figure 4(b)-(c) illustrates the expected charge distribution and band diagram of the Al₂O₃/Al$_x$Ga$_{1-x}$N/Al$_y$Ga$_{1-y}$N double-interface structure under flat-band conditions. Based on the energy band diagram, the flat-band voltage ($V_{FB}$) is derived as:

$$V_{FBi} = \left(\frac{\varphi_b - \Delta E_C - \varphi_s}{q}\right) - F_{ox}t_{oxi} - (P_{T.b} - P_{sp.ch} - \frac{N_D}{2\varepsilon_b}t_b)t_b$$

(1)

where $\varphi_b$ is the Schottky barrier height, $\Delta E_C$ is the conduction band offset, $\varphi_s$ is the energy difference between the Fermi level and conduction band energy in the channel layer, $q$ is the electron charge value, $F_{ox}$ is the oxide field intensity, $t_b$ is AlGaN barrier thickness, $V_{FBi}$ is flat-band voltage corresponding to each Al₂O₃ thickness ($t_{oxi}$), and $P_{T.b}$ and $P_{sp.ch}$ is total polarization charge density of barrier and the spontaneous polarization charge density of channel, respectively. This equation suggests that $V_{FBi}$ is a linear function of $t_{oxi}$ where the slope is $F_{ox}$, while all other parameters are a constant for the linear function. The flat-band voltage was extracted by differentiating the measured C-V curves for each oxide thickness. Figure 5 presents the measured C-V characteristics, where the 13 nm Al₂O₃ C-V data for the 85/68 sample appears as an outlier, located in a highly resistive region of the sample. Across all samples, the C-V measurements exhibited minimal hysteresis, with negligible voltage variation under double-sweep conditions.

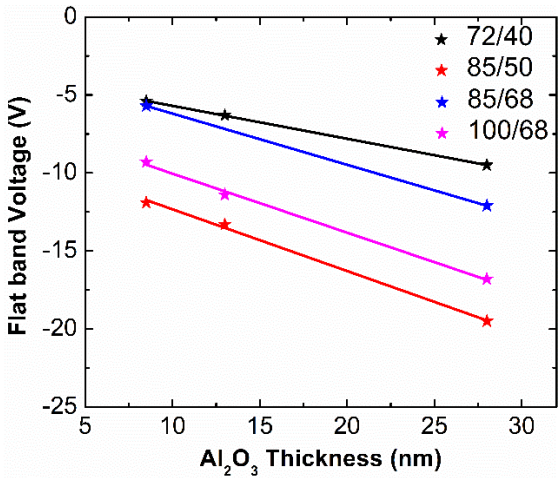

Figure 6. Flat-band voltage vs Al₂O₃ thickness. The slope of each linear functions indicates the internal oxide field intensity.

Table 2. Extracted internal oxide field and fixed interface charge density

| | $N_{D.b}^{+}$ [cm$^{-3}$] | $P_{sp.ch}$ [x10$^{13}$ cm$^{-2}$] | $F_{ox}$ [MV/cm] | $\sigma_s$ [x10$^{13}$ cm$^{-2}$] | $\sigma_{net}$ [x10$^{13}$ cm$^{-2}$] |
|---|---|---|---|---|---|
| 72/40 | 2x10$^{18}$ | -3.11 | 3.74 | 4.37 | 1.86 |
| 85/50 | 3x10$^{18}$ | -3.43 | 3.96 | 4.50 | 1.97 |
| 85/68 | 4x10$^{18}$ | -4.02 | 3.28 | 4.45 | 1.63 |
| 100/68 | UID | -4.02 | 3.78 | 5.90 | 1.88 |

The $F_{ox}$ for each sample was estimated from the slope of the linear fits in Figure 6. based on the derived equation (1). Additionally, the fixed interface charge density was calculated from the correlation of field and charge (Figure 4(b)), which is given by:

$$\sigma_s = F_{ox}\epsilon_{ox} - N_{D.b}^{+}t_b + P_{sp.ch} \qquad (2)$$

Where $\sigma_s$ is the fixed interface charge density, $\epsilon_{ox}$ is oxide permittivity, and $N_{D.b}^{+}$ is doping concentration of barrier layer. The evaluated $F_{ox}$ and $\sigma_s$ at the interface for different Al composition on barrier and channel layers are summarized in Table 2. This study confirms the presence of positive fixed charges at the interfaces between PEALD Al$_2$O$_3$ and UWBG AlGaN, with the $\sigma_s$ being at least 2× higher than that of the Al$_2$O$_3$/GaN interface, leading to higher oxide fields ($F_{ox}$) within the oxide [30, 33].

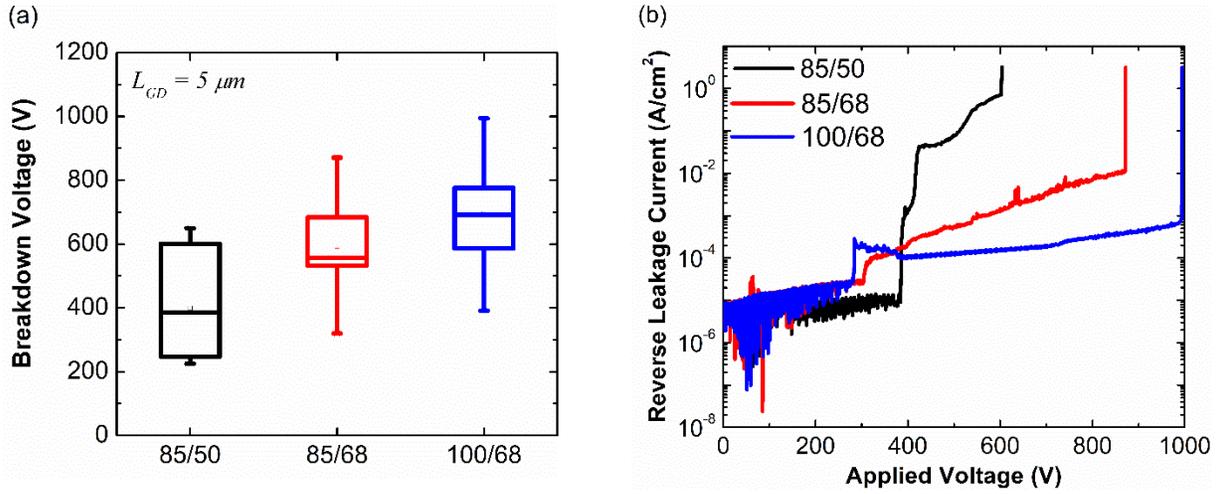

Figure 7. (a) Breakdown voltage trend for each sample with gate-drain spacing ($L_{GD}$) of 5 $\mu m$, (b) two-terminal breakdown measurements

Table 3. Effective peak electric field and average breakdown field for gate-drain spacing of 5 $\mu m$

|  | $F_{ox}$ [MV/cm] | $F_{eff,OX}$ [MV/cm] | $F_{Lat,Av}$ [MV/cm] |
|---|---|---|---|
| 85/50 | 3.96 | > 4.04 | 0.79 |
| 85/68 | 3.28 | > 3.47 | 1.14 |
| 100/68 | 3.78 | > 4.27 | 1.99 |

Two-terminal gate-to-drain breakdown characteristics are presented in Figure 7(b), with measurements conducted using Keysight B1505A. Breakdown in this case was defined as the voltage at which the current density reached $1 \times 10^{-3}$ A/cm$^2$. Regardless of Al composition, no systematic dependence on Al$_2$O$_3$ thickness was observed. Despite the presence of higher fixed interface charge density, Al$_2$O$_3$-integrated UWBG AlGaN HEMT structures demonstrated the capability to support high breakdown voltage. For the 100/68 sample, the breakdown voltage reached ~ 1 kV ($L_{GD}$ = 5 $\mu$m) for a sheet charge density of $1.27 \times 10^{13}$ cm$^{-2}$ without employing any field management techniques such as field plates and passivation (Figure 7(a)). These are significantly higher than what would be achieved in GaN-channel devices without the use of field termination and show the potential of UWBG materials for future high-breakdown III-Nitride technology. Under these breakdown conditions, the x-direction peak electric field is obviously higher than average lateral breakdown field ($F_{Lat,Av} > \frac{V_{BR}}{L_{GD}}$). The effective peak electric field ($F_{eff,OX}$) at gate edge in the oxide under breakdown condition is given by:

$$F_{eff} = \sqrt{F_x^2 + F_y^2} > \sqrt{(\frac{V_{BR}}{L_{GD}})^2 + F_{ox}^2} \qquad (3)$$

Following equation (3), the estimated $F_{eff}$, and average breakdown field are summarized in Table 3. The estimated oxide field due to fixed charges was ~3.96 MV/cm. Although the fixed positive interface charge density was higher than that of Al$_2$O$_3$/GaN, PEALD Al$_2$O$_3$-integrated UWBG AlGaN lateral test structures with AlN/68% AlGaN structure demonstrated 1 kV breakdown voltage with a sheet charge density of $1.27 \times 10^{13}$ cm$^{-2}$ even without any field management techniques like field plates. The extracted effective peak electric field and average breakdown field were > 4.27 MV/cm and 1.99 MV/cm, respectively. All samples exhibited low leakage currents (~5 × 10$^{-7}$ A/cm$^2$ at < 200 V or until breakdown).

In conclusion, interfacial and breakdown characteristics of Al$_2$O$_3$ on UWBG AlGaN HEMT structures were investigated. This work shows the importance of understanding and controlling interface fixed charges at the oxide interface, since these fixed charges can greatly impact the total effective field within the oxide. Lateral test structures showed remarkable breakdown behavior, with lateral average fields in excess of 2 MV/cm across spacings as large as 5 $\mu m$, which are significantly higher than fields achieved in conventional GaN-channel transistors. Further implementation of optimized field management strategies will provide even greater improvements over incumbent technologies. This study therefore highlights the potential of $Al_2O_3$-integrated UWBG AlGaN structures to surpass conventional GaN technology in terms of breakdown field and scaling of device dimensions for future power and RF applications.

This work was funded by ARO DEVCOM under Grant No. W911NF2220163 (UWBG RF Center, program manager Dr. Tom Oder). This article has been authored by an employee of National Technology & Engineering Solutions of Sandia, LLC under Contract No. DE-NA0003525 with the U.S. Department of Energy (DOE). The employee owns all right, title and interest in and to the article and is solely responsible for its contents. The United States Government retains and the publisher, by accepting the article for publication, acknowledges that the United States Government retains a non-exclusive, paid-up, irrevocable, world-wide license to publish or reproduce the published form of this article or allow others to do so, for United States Government purposes. The DOE will provide public access to these results of federally sponsored research in accordance with the DOE Public Access Plan https://www.energy.gov/downloads/doe-public-access-plan

**Author Declarations**

**Conflict of Interest**

The authors have no conflicts to disclose.

**Data Availability**

The data that support the findings of this study are available within the article.

## References

[1] M. E. Coltrin, A. G. Baca, and R. J. Kaplar, "Analysis of 2D Transport and Performance Characteristics for Lateral Power Devices Based on AlGaN Alloys," *ECS J. Solid State Sci. Technol.*, vol. 6, no. 11, p. S3114, Oct. 2017, doi: 10.1149/2.0241711jss.

[2] M. E. Coltrin and R. J. Kaplar, "Transport and breakdown analysis for improved figure-of-merit for AlGaN power devices," *Journal of Applied Physics*, vol. 121, no. 5, p. 055706, Feb. 2017, doi: 10.1063/1.4975346.

[3] R. J. Kaplar et al., "Review—Ultra-Wide-Bandgap AlGaN Power Electronic Devices," ECS J. Solid State Sci. Technol., vol. 6, no. 2, p. Q3061, Dec. 2016, doi: 10.1149/2.0111702jss.

[4] A. G. Baca, A. M. Armstrong, B. A. Klein, A. A. Allerman, E. A. Douglas, and R. J. Kaplar, "Al-rich AlGaN based transistors," Journal of Vacuum Science & Technology A: Vacuum, Surfaces, and Films, vol. 38, no. 2, p. 020803, Mar. 2020, doi: 10.1116/1.5129803.

[5] T. L. Chu and R. W. Kelm, "The Preparation and Properties of Aluminum Nitride Films," J. Electrochem. Soc., vol. 122, no. 7, p. 995, Jul. 1975, doi: 10.1149/1.2134385.

[6] J. L. Hudgins, G. S. Simin, E. Santi, and M. A. Khan, "An assessment of wide bandgap semiconductors for power devices," IEEE Transactions on Power Electronics, vol. 18, no. 3, pp. 907–914, May 2003, doi: 10.1109/TPEL.2003.810840.

[7] Y. Zhu et al., "Heterostructure and interfacial engineering for low-resistance contacts to ultra-wide bandgap AlGaN," Applied Physics Letters, vol. 126, no. 6, p. 062107, Feb. 2025, doi: 10.1063/5.0230220.

[8] P. Kordoš, G. Heidelberger, J. Bernát, A. Fox, M. Marso, and H. Lüth, "High-power SiO2∕AlGaN∕GaN metal-oxide-semiconductor heterostructure field-effect transistors," Applied Physics Letters, vol. 87, no. 14, p. 143501, Sep. 2005, doi: 10.1063/1.2058206.

[9] N. Maeda et al., "Systematic Study of Insulator Deposition Effect (Si3N4, SiO2, AlN, and Al2O3) on Electrical Properties in AlGaN/GaN Heterostructures," Jpn. J. Appl. Phys., vol. 46, no. 2R, p. 547, Feb. 2007, doi: 10.1143/JJAP.46.547

[10] C. J. Kirkpatrick, B. Lee, R. Suri, X. Yang, and V. Misra, "Atomic Layer Deposition of \hboxSiO_2 for AlGaN/GaN MOS-HFETs," IEEE Electron Device Letters, vol. 33, no. 9, pp. 1240–1242, Sep. 2012, doi: 10.1109/LED.2012.2203782.

[11] H. Kambayashi et al., "High Quality SiO2/Al2O3 Gate Stack for GaN Metal–Oxide–Semiconductor Field-Effect Transistor," Jpn. J. Appl. Phys., vol. 52, no. 4S, p. 04CF09, Mar. 2013, doi: 10.7567/JJAP.52.04CF09.

[12] J.-G. Lee, H.-S. Kim, K.-S. Seo, C.-H. Cho, and H.-Y. Cha, "High quality PECVD SiO2 process for recessed MOS-gate of AlGaN/GaN-on-Si metal–oxide–semiconductor heterostructure field-effect transistors," Solid-State Electronics, vol. 122, pp. 32–36, Aug. 2016, doi: 10.1016/j.sse.2016.04.016.

[13] Z. Liu et al., "Investigation of the interface between LPCVD-SiNx gate dielectric and III-nitride for AlGaN/GaN MIS-HEMTs," Journal of Vacuum Science & Technology B, vol. 34, no. 4, p. 041202, Mar. 2016, doi: 10.1116/1.4944662.

[14] Z. Zhang et al., "16.8 A/600 V AlGaN/GaN MIS-HEMTs employing LPCVD-Si3N4 as gate insulator," Electronics Letters, vol. 51, no. 15, pp. 1201–1203, 2015, doi: 10.1049/el.2015.1018.

[15] X. Lu, K. Yu, H. Jiang, A. Zhang, and K. M. Lau, "Study of Interface Traps in AlGaN/GaN MISHEMTs Using LPCVD SiNx as Gate Dielectric," IEEE Transactions on Electron Devices, vol. 64, no. 3, pp. 824–831, Mar. 2017, doi: 10.1109/TED.2017.2654358.

[16] X.-Y. Liu et al., "AlGaN/GaN MISHEMTs with AlN gate dielectric grown by thermal ALD technique," Nanoscale Res Lett, vol. 10, no. 1, p. 109, Mar. 2015, doi: 10.1186/s11671-015-0802-x.

[17] J.-J. Zhu et al., "Improved Interface and Transport Properties of AlGaN/GaN MIS-HEMTs With PEALD-Grown AlN Gate Dielectric," IEEE Transactions on Electron Devices, vol. 62, no. 2, pp. 512–518, Feb. 2015, doi: 10.1109/TED.2014.2377781.

[18] Y. Lu, S. Yang, Q. Jiang, Z. Tang, B. Li, and K. J. Chen, "Characterization of VT-instability in enhancement-mode Al2O3-AlGaN/GaN MIS-HEMTs," physica status solidi c, vol. 10, no. 11, pp. 1397–1400, 2013, doi: 10.1002/pssc.201300270.

[19] Z. Wang et al., "Thin-barrier enhancement-mode AlGaN/GaN MIS-HEMT using ALD Al2O3 as gate insulator*," J. Semicond., vol. 36, no. 9, p. 094004, Sep. 2015, doi: 10.1088/1674-4926/36/9/094004.

[20] H.-C. Wang et al., "AlGaN/GaN MIS-HEMTs With High Quality ALD-Al2O3 Gate Dielectric Using Water and Remote Oxygen Plasma As Oxidants," IEEE Journal of the Electron Devices Society, vol. 6, pp. 110–115, 2018, doi: 10.1109/JEDS.2017.2779172.

[21] T. Kubo, J. J. Freedsman, Y. Iwata, and T. Egawa, "Electrical properties of GaN-based metal–insulator–semiconductor structures with Al2O3 deposited by atomic layer deposition using water and ozone as the oxygen precursors," Semicond. Sci. Technol., vol. 29, no. 4, p. 045004, Feb. 2014, doi: 10.1088/0268-1242/29/4/045004.

[22] Myoung-Jin Kang, "A study on the dielectric layers for high-power AlGaN/GaN devices," Thesis,

Seoul National University 2020. Accessed: Feb. 27, 2025. [Online]. Available: https://s-space.snu.ac.kr/handle/10371/168028

[23] R. Lossy, H. Gargouri, M. Arens, and J. Würfl, "Gallium nitride MIS-HEMT using atomic layer deposited Al2O3 as gate dielectric," Journal of Vacuum Science & Technology A, vol. 31, no. 1, p. 01A140, Dec. 2012, doi: 10.1116/1.4771655.

[24] T.-E. Hsieh et al., "Gate Recessed Quasi-Normally OFF Al2O3/AlGaN/GaN MIS-HEMT With Low Threshold Voltage Hysteresis Using PEALD AlN Interfacial Passivation Layer," IEEE Electron Device Letters, vol. 35, no. 7, pp. 732–734, Jul. 2014, doi: 10.1109/LED.2014.2321003.

[25] T. Kubo, M. Miyoshi, and T. Egawa, "Post-deposition annealing effects on the insulator/semiconductor interfaces of Al2O3/AlGaN/GaN structures on Si substrates," Semicond. Sci. Technol., vol. 32, no. 6, p. 065012, May 2017, doi: 10.1088/1361-6641/aa6c09.

[26] E. Schilirò et al., "Early Growth Stages of Aluminum Oxide (Al2O3) Insulating Layers by Thermal- and Plasma-Enhanced Atomic Layer Deposition on AlGaN/GaN Heterostructures," ACS Appl. Electron. Mater., vol. 4, no. 1, pp. 406–415, Jan. 2022, doi: 10.1021/acsaelm.1c01059.

[27] S. Huang, S. Yang, J. Roberts, and K. J. Chen, "Characterization of Vth-instability in Al2O3/GaN/AlGaN/GaN MIS-HEMTs by quasi-static C-V measurement," physica status solidi c, vol. 9, no. 3–4, pp. 923–926, 2012, doi: 10.1002/pssc.201100302.

[28] R. Meunier, A. Torres, E. Morvan, M. Charles, P. Gaud, and F. Morancho, "AlGaN/GaN MIS-HEMT gate structure improvement using Al2O3 deposited by plasma-enhanced ALD," Microelectronic Engineering, vol. 109, pp. 378–380, Sep. 2013, doi: 10.1016/j.mee.2013.04.020.

[29] R. Meunier, A. Torres, M. Charles, E. Morvan, M. Plissonier, and F. Morancho, "AlGaN/GaN MIS-HEMT Gate Structure Improvement Using Al2O3 Deposited by PEALD and BCl3 Gate Recess Etching," ECS Trans., vol. 58, no. 4, p. 269, Aug. 2013, doi: 10.1149/05804.0269ecst.

[30] T.-H. Hung, S. Krishnamoorthy, M. Esposto, D. Neelim Nath, P. Sung Park, and S. Rajan, "Interface charge engineering at atomic layer deposited dielectric/III-nitride interfaces," Applied Physics Letters, vol. 102, no. 7, p. 072105, Feb. 2013, doi: 10.1063/1.4793483.

[31] H. Xue et al., "Al 0.75 Ga 0.25 N/Al 0.6 Ga 0.4 N heterojunction field effect transistor with f T of 40 GHz," Appl. Phys. Express, vol. 12, no. 6, p. 066502, Jun. 2019, doi: 10.7567/1882-0786/ab1cf9.

[32] X. Hu et al., "Doped Barrier Al0.65Ga0.35N/Al0.40Ga0.60N MOSHFET With SiO2 Gate-Insulator and Zr-Based Ohmic Contacts," IEEE Electron Device Letters, vol. 39, no. 10, pp. 1568–1571, Oct. 2018, doi: 10.1109/LED.2018.2866027.

[33] H. Zhang et al., "Presence of High Density Positive Fixed Charges at ALD–Al2O3/GaN (cap) Interface for Efficient Recovery of 2-DEG in Ultrathin-Barrier AlGaN/GaN Heterostructure," physica status solidi (b), vol. 261, no. 11, p. 2300555, 2024, doi: 10.1002/pssb.202300555.

[34] Zhang, Z., Jackson, C. M., Arehart, A. R., McSkimming, B., Speck, J. S., & Ringel, S. A. (2014). Direct Determination of Energy Band Alignments of Ni/Al2O3/GaN MOS Structures Using Internal Photoemission Spectroscopy. Journal of Electronic Materials, 43(4), 828–832. https://doi.org/10.1007/s11664-013-2942-z

[35] Di Lecce, V., Krishnamoorthy, S., Esposto, M., Hung, T.-H., Chini, A., & Rajan, S. (2012). Metal-oxide barrier extraction by Fowler-Nordheim tunnelling onset in Al2O3-on-GaN MOS diodes. Electronics Letters, 48(6), 347–348. https://doi.org/10.1049/el.2011.4046

[36] Hashizume, T., Kaneki, S., Oyobiki, T., Ando, Y., Sasaki, S., & Nishiguchi, K. (2018). Effects of postmetallization annealing on interface properties of Al2O3/GaN structures. Applied Physics Express, 11(12), 124102. https://doi.org/10.7567/APEX.11.124102

[37] Hori, Y., Mizue, C., & Hashizume, T. (2010). Process Conditions for Improvement of Electrical Properties of Al2O3/n-GaN Structures Prepared by Atomic Layer Deposition. Japanese Journal of Applied Physics, 49(8R), 080201. https://doi.org/10.1143/JJAP.49.080201

[38] Ganguly, S., Verma, J., Li, G., Zimmermann, T., Xing, H., & Jena, D. (2011). Presence and origin of interface charges at atomic-layer deposited Al2O3/III-nitride heterojunctions. Applied Physics Letters, 99(19), 193504. https://doi.org/10.1063/1.3658450

[39] Ostermaier, C., Lee, H.-C., Hyun, S.-Y., Ahn, S.-I., Kim, K.-W., Cho, H.-I., Ha, J.-B., & Lee, J.-H. (2008). Interface characterization of ALD deposited Al2O3 on GaN by CV method. Physica Status Solidi c, 5(6), 1992–1994. https://doi.org/10.1002/pssc.200778663